\documentclass[12pt]{iopart}
\usepackage{iopams}
\usepackage{graphicx}
\usepackage{color}

\begin{document}
\noindent
Accepted for publication at: \\Journal of Physics B: Atomic, Molecular and Optical Physics
Special Issue: Modern Applications of Trapped Ions
{\,\,}

\title{High-accuracy Penning trap mass measurements with stored and
cooled exotic ions}
\author{K. Blaum$^1$, Sz. Nagy$^1$ and G. Werth$^2$}
\address{$ˆ1$ Max-Planck-Institut f\"ur Kernphysik, D-69117 Heidelberg, Germany}
\address{$ˆ2$ Institut f\"ur Physik, Johannes Gutenberg-Universit\"at, D-55099 Mainz, Germany}
\ead{klaus.blaum@mpi-hd.mpg.de}
\begin{abstract}
The technique of Penning trap mass spectrometry is briefly reviewed
particularly in view of precision experiments on unstable nuclei,
performed at different facilities worldwide. Selected examples of
recent results emphasize the importance of high-precision mass
measurements in various fields of physics.
\end{abstract}
\submitto{\jpb}
\maketitle

\section{Introduction}
The mass is a unique, fundamental property of an atom. Mass values
provide access to nuclear and atomic binding energies, serve for
comparison with nuclear models as well as for test of fundamental
interactions like quantum-electrodynamics or weak interaction
\cite{Bollen2001,Lunney2003,Blaum2006,Franzke2008}. The required precision
depends on the problem: while for particle identification in physics
and chemistry a relative mass uncertainty $\delta m/m$ of $10^{-5}$ is
sufficient, tests of bound state quantum electrodynamics in heavy
ions require $\delta m /m < 10^{-11}$. Table \ref{tab:applications}
gives examples of the required mass uncertainty for different
applications.

\begin{table} [htbp]
\caption{\label{tab:applications}Some applications of mass
determinations and typically required relative mass uncertainty
$\delta m/m$.}
\begin{indented}
\item[]\begin{tabular}{@{}ll}
\br
application of mass determination&$\delta m/m$\\
\mr
Identification of atoms and molecules&$<10^{-5}$\\
Nuclear structure&$<10^{-6}$\\
Astrophysics&$<10^{-6}$\\
Weak interaction studies&$<10^{-8}$\\
Metrology-Fundamental constants&$<10^{-9}$\\
QED in highly charged ions&$<10^{-11}$\\
\br
\end{tabular}
\end{indented}
\end{table}

\begin{figure} [htbp]
    \centering
        \includegraphics[width=0.50\textwidth]{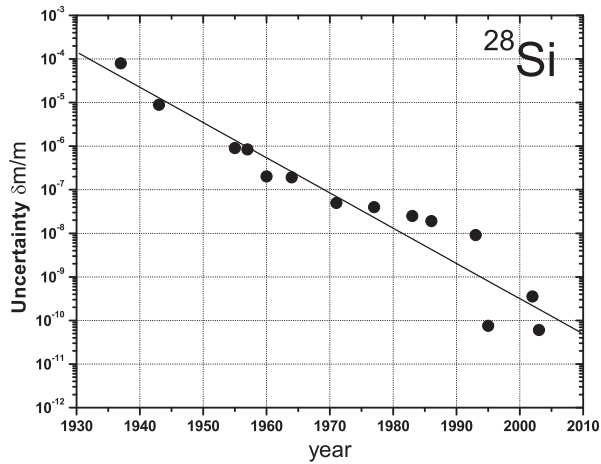}
    \caption{Decrease of mass uncertainties in time for $^{28}$Si using various techniques \cite{Audi2006}.}
    \label{fig:mass-uncertainties}
\end{figure}

In the history of mass spectrometry the precision of atomic mass
determination has shown a constant improvement of about an order of
magnitude every decade \cite{Audi2006}. Figure
\ref{fig:mass-uncertainties} shows as an example the decreasing
uncertainty of the mass of $^{28}$Si in time using various
techniques. The introduction of devices that employed frequency
measurements in the 1950s \cite{Smith54}, in particular the introduction of
Penning traps in the 1980s, was a breakthrough in mass spectrometry which lead to
a dramatic decrease in the mass uncertainty.

In this contribution we will focus on mass determinations of
\emph{exotic} ions such as short lived radionuclides, and
highly charged ions. We will briefly introduce the technique of Penning traps, describe the basic features of facilities for mass determinations of radionuclides and
discuss some of the applications of precise mass determinations.

\section{Principle of Penning trap mass spectrometry}

Mass determination in Penning traps relies on the fact that the
ratio of cyclotron frequencies $\nu_{{c}} = qeB/(2\pi m)$ for two
ion species in the same magnetic field $B$ gives directly their mass
ratio:

\begin{equation}
\frac{\nu_{{c}}(1)}{\nu_{{c}}(2)} = \frac{q(1)}{q(2)} \cdot
\frac{m(2)}{m(1)}.
\end{equation}

\begin{figure} [htbp]
    \centering
        \includegraphics[width=0.60\textwidth]{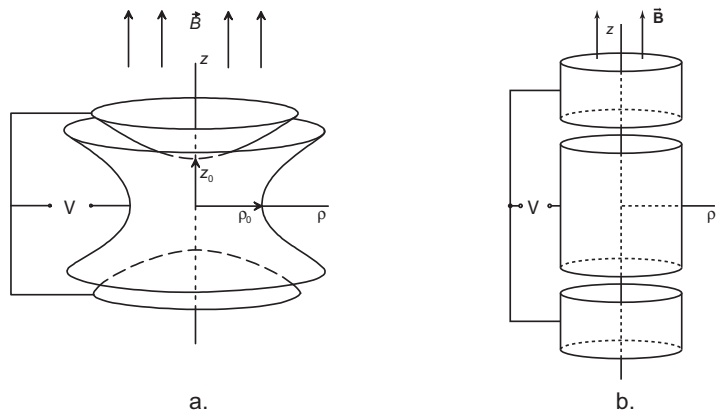}
    \caption{(a) Hyperbolical and (b) cylindrical Penning trap.}
    \label{fig:Penning-traps}
\end{figure}

\begin{figure} [htbp]
    \centering
        \includegraphics[width=0.50\textwidth]{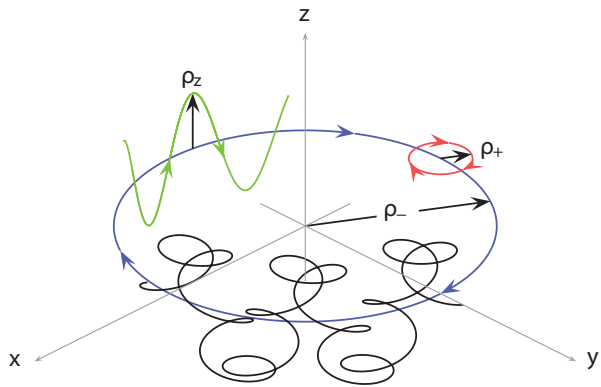}
    \caption{Motion of a single charged particle in a Penning trap.
    $\rho_+$, $\rho_z$, and $\rho_-$ are the radii of the cyclotron, axial,
    and magnetron motions, respectively. The black solid line represents a
    superposition of the three eigenmotions. The frequencies
    and amplitudes are not to scale.}
    \label{fig:ion-motion}
\end{figure}

\noindent $q$ and $m$ are the charge state and mass, respectively,
of each ion. If $^{12}$C as atomic mass standard is taken as
reference, the mass of the unknown species is obtained in atomic
mass units \cite{Blaum2002}. The ions are confined for extended
periods of time in a small volume in space by a strong magnetic
field and a weak electric quadrupole potential created by
hyperbolically shaped electrodes (figure \ref{fig:Penning-traps}a)
of the Penning trap. Alternatively cylindrical electrodes may be
used (figure \ref{fig:Penning-traps}b) which are easier to
manufacture and to align. Near the center of the cylindrical trap
the potential can be well approximated by a quadrupole potential.
Higher order contributions can be partially reduced by guard
electrodes placed between the central ring and the outer endcap
electrodes \cite{Gabrielse84}.
\noindent The quadrupole potential is given by

\begin{equation}\Phi = \frac{V}{d^{2}} \left(\rho^{2} - 2z^{2}\right),
\end{equation}

\noindent where $\rho^{2}=x^{2} + y^{2}$ and $2d^{2} = z_{0}^{2} +
\rho_{0}^{2}/2$ describe the dimensions of the trap (with $2\rho_0$
and $2z_0$ being the inner ring diameter and the closest distance
between the endcap electrodes, respectively, see figure
\ref{fig:Penning-traps}). $V$ is the applied potential difference
between the ring and endcaps. A single particle of charge $q$ and mass $m$ oscillates in the axial direction with frequency
\begin{equation}
\nu_{{z}} = \frac{1}{2\pi}\sqrt{\frac{qV}{md^{2}}}.
\label{eq:axial-frequency}
\end{equation}
\noindent In the radial direction we have a superposition of two motions with frequencies
\begin{equation}
\nu_{+} = \frac{\nu_{{c}}}{2} + \left(\frac{\nu_{{c}}^{2}}{4} -
\frac{\nu_{{z}}^{2}}{2}\right)^{1/2}
\label{eq:perturbed-cyclotron-frequency}
\end{equation}
\noindent and
\begin{equation}
\nu_{-} = \frac{\nu_{{c}}}{2} - \left(\frac{\nu_{{c}}^{2}}{4} -
\frac{\nu_{{z}}^{2}}{2}\right)^{1/2}. \label{eq:magnetron-frequency}
\end{equation}
\noindent $\nu_{+}$ is called the reduced cyclotron frequency and
$\nu_{-}$ the magnetron frequency. Typically $\nu_{+}$ is in the MHz
range, $\nu_{-}$ is a few kHz and $\nu_{z}$ several 10\,kHz. The
superposition of these three eigenfrequencies leads to a motion as
illustrated in figure \ref{fig:ion-motion}. For more details on the
principles of Penning traps see \cite{Brown1986,Major2005}.
\noindent The cyclotron frequency as required for mass determination
is not an eigenfrequency of the particles motion. It can, however,
be obtained by the sum of $\nu_{+}$ and $\nu_{-}$. Deviations of the
potential from the ideal quadrupole shape shift the eigenfrequencies
and may lead to errors in mass determination. Brown and Gabrielse
have shown that the relation
\begin{equation}\nu_{{c}}^{2} = \nu_{+}^{2} + \nu_{-}^{2} +
\nu_{{z}}^{2},
\end{equation}
\noindent known as the \emph{Brown-Gabrielse invariance theorem}, is
independent on potential perturbations to first order
\cite{Brown1982,Gabrielse2008}.
The detection of the motional frequencies can be performed in
different ways. In the \emph{non-destructive detection technique} the particle
remains in the trap and does not get lost in the detection process.
Repeated measurements can be carried out with relatively long
observation times by picking up the voltage induced in the trap
electrodes by the oscillating charged particles employing sensitive
electronics with a very high quality factor $Q$. A Fourier transform
of the induced voltage reveals the motional frequency. Another
advantage of this method is that thanks to its high sensitivity it
requires a very small number of simultaneously stored ions.
Measurements on single particles have been demonstrated which
eliminate potential frequency shifts arising from the Coulomb field
of simultaneously confined ions. The trapped ion can be cooled to
low energies by keeping its oscillation in resonance with a tank
circuit attached to the trap and kept at liquid helium temperature
(\emph{resistive cooling}). Then the oscillation amplitude is of the
order of a few microns and residual potential imperfections or
magnetic field inhomogeneities seen by the ion play only very small
role. In fact the highest precision in mass spectrometry has been
obtained using this techniques
\cite{Bradley1999,Dyck1999,Gabrielse1999,Rainville2004,Shi2005}.

\begin{figure} [htbp]
    \centering
        \includegraphics[width=0.50\textwidth]{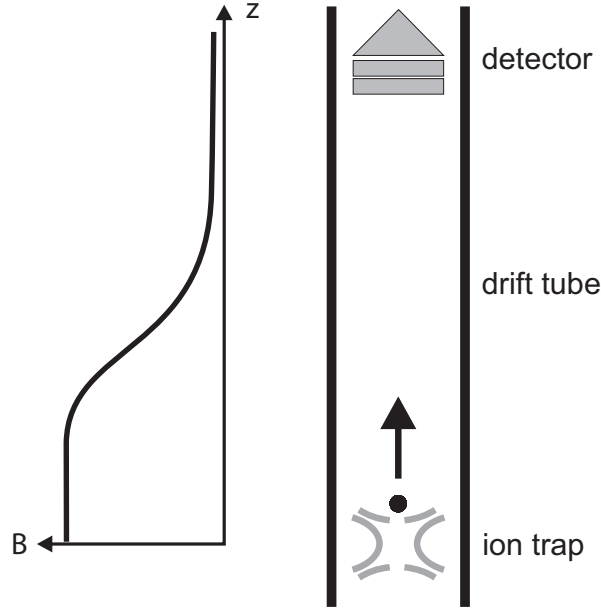}
    \caption{Principle of the time-of-flight method for cyclotron resonance detection.}
    \label{fig:TOF-principle}
\end{figure}

Because of the time required to cool the trapped ions and to
accumulate sufficient signal strength (transients) for a good
signal-to-noise ratio in the Fourier spectrometer, this
non-destructive detection method is not suited for investigation of
very short-lived radionuclides with half-lives below one second
\cite{Weber2005}. Instead a \emph{destructive detection method} for detecting
the cyclotron resonance has been developed \cite{Graeff1980} and is
known as the time-of-flight ion-cyclotron-resonance technique
(ToF-ICR). ToF-ICR is a destructive technique in the sense that the
trapped ion is lost in the detection process and the trap has to be
reloaded in each cycle. The technique involves manipulating the ions
eigenmotions and probing the cyclotron frequency using an external
radiofrequency field and measuring the flight time of the ions
ejected from the trap to a detector. Since we focus on experiments
with unstable nuclei we shall describe this technique in more
detail.
\begin{figure} [htbp]
    \centering
        \includegraphics[width=0.70\textwidth]{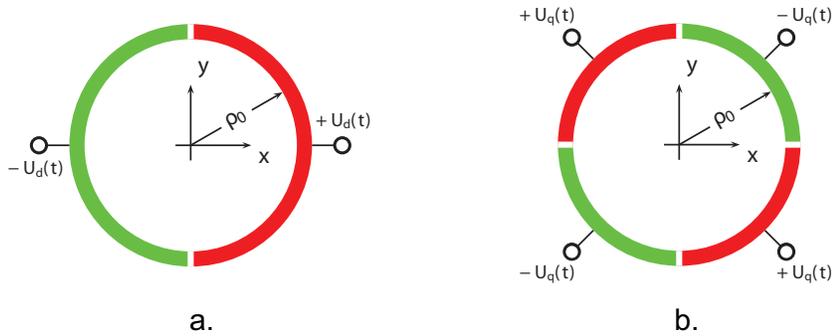}
    \caption{Segmented ring electrode for application of (a) a dipole field
    (b) a quadrupole field with amplitudes $U_{{d}}$ and $U_{{q}}$, respectively,
    for excitation of the radial motions.}
    \label{fig:segmented-ring-electrode}
\end{figure}
The ion motion can be manipulated by external radiofrequency
fields: With dipolar excitation (see figure
\ref{fig:segmented-ring-electrode}a) typically the amplitude of an
eigenmotion is enhanced, while quadrupolar fields (see figure
\ref{fig:segmented-ring-electrode}b) are used to couple two
eigenmotions. Within the ToF-ICR technique the trapped ions are
first prepared by a phase-locked dipolar excitation to a defined
radius of the magnetron motion \cite{Blaum2003a}. At this stage the
orbital magnetic moment $\mu$ and the corresponding energy $E=
\vec{\mu} \vec{B}$ are small. By applying a resonant quadrupolar
radiofrequency excitation signal with properly chosen amplitude and
duration on the segmented ring electrode, the radial motions couple,
and the magnetron motion is completely converted into the modified
cyclotron motion leading to a large increase in radial energy. The
excited ions are ejected from the trap towards an ion counting
detector, see figure \ref{fig:TOF-principle}. While traveling through
the fringe field of the magnet the ions get accelerated due to the
gradient force:
\begin{equation}
\vec{F}= -\vec{\mu} (\vec{\nabla} \vec{B})= - \frac{E_r}{B}\frac{\partial B}{\partial z} \hat{z}.
\end{equation}
The force is proportional to the orbital magnetic moment and thus to
the radial energy $E_r$. The magnetic moment and the radial energy
of the ions is the largest at resonance, when the frequency of the
exciting field equals the true cyclotron frequency ($\nu_c$) which
leads to a reduction in the time of flight from the trap to the
detector. The total time of flight from the trap center ($z=0$) to
the detector ($z=z_1$) for a given radial energy $E_r$ is given by:
\begin{equation}
T_{tot}(\omega_q)=\int^{z_1}_{0} \sqrt{\frac{m}{2(E_0-qU(z)-\mu(\omega_q)B(z)}} \,\, dz,
\end{equation}
where $\omega_q$ is the frequency of the exciting field, $E_0$
denotes the initial axial energy and $U(z)$ and $B(z)$ the electric
and magnetic potential difference, respectively.

\begin{figure} [htbp]
    \centering
        \includegraphics[width=0.70\textwidth]{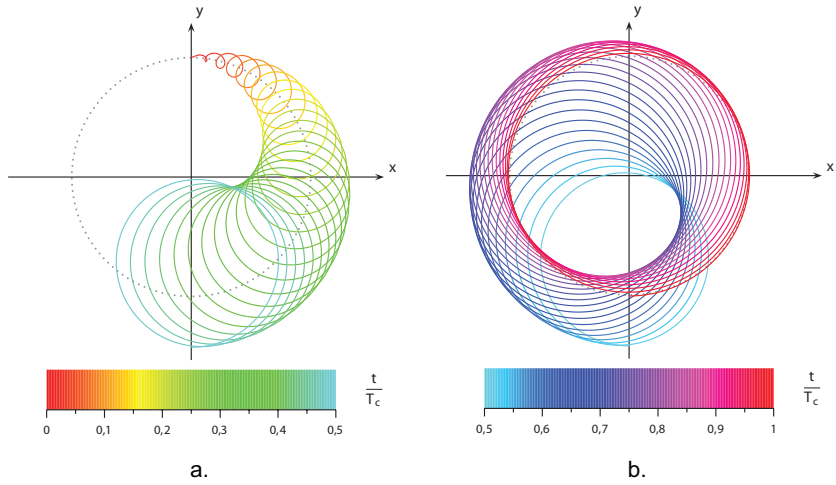}
    \caption{Calculated conversion of a pure magnetron
    motion into a pure cyclotron motion by a radial quadrupolar rf
    field at frequency $\nu_{{c}} = \nu_{+} + \nu_{-}$.
    The time for one full conversion is $T_c$.
    Parts (a) and (b) show the first and second half of the conversion, respectively.}
    \label{fig:motion-conversion}
\end{figure}

\noindent When the exciting quadrupolar field is applied with
constant amplitude $U_q$ for a fixed time interval $T_{{rf}}$, the
resulting energy gain $E_{{r}}$ is \cite{Koenig1995}
\begin{equation}
E_{{r}} = \frac{\sin^{2}(\omega_{{b}} T_{{rf}})}{\omega_{{b}}^{2}}
\label{eq:energy-gain}
\end{equation}
\noindent with
\begin{equation}
\omega_{{b}} = \frac{1}{2} \sqrt{(\omega_{{rf}} - \omega_{{c}})^{2} + (\omega_{{conv}}/2)^{2}}.
\label{eq:frequency-b}
\end{equation}
\noindent The conversion frequency $\omega_{{conv}}$ is given by
\begin{equation}
\omega_{{conv}} = \frac{U_{{q}}}{r_{0}^{2}} \cdot \frac{1}{4B},
\label{eq:conversion-frequency}
\end{equation}
where $U_q$ corresponds to the maximum potential of the quadrupole radiofrequency field
measured on a circle with radius $r_0$.
Hence, the effect of the coupling is a periodic conversion of the
perturbed cyclotron and the magnetron oscillations \cite{Koenig1995}
as shown in figure \ref{fig:motion-conversion}.

\begin{figure} [htbp]
    \centering
        \includegraphics[width=0.60\textwidth]{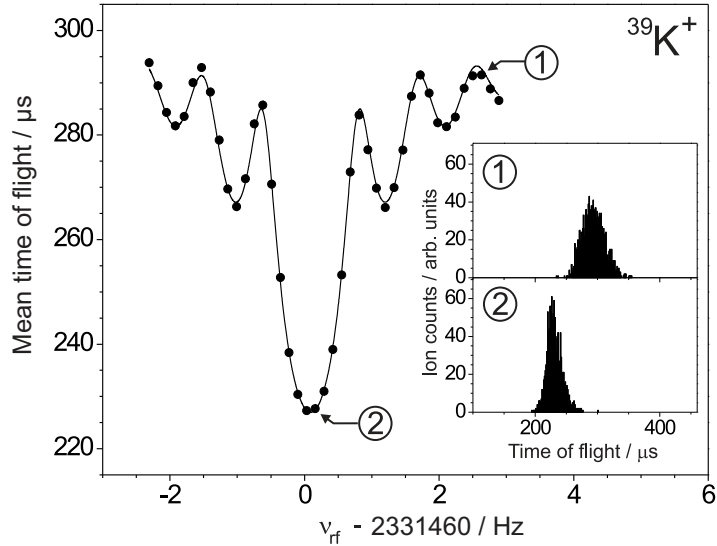}
    \caption{Cyclotron resonance of $^{39}$K$^{+}$. The averaged
    data are fitted to the theoretical line shape \cite{Koenig1995} for an excitation time of
    ${T}_{{rf}} = 900$ ms. The insert shows two time-of-flight
    spectra after accumulation of many single ion events.
    (1) and (2) are off and on resonance, respectively.}
    \label{fig:cyclotron-resonance}
\end{figure}

In the ToF-ICR technique the energy gain is directly transformed
into a change in time of flight, and for optimum energy conversion
into the cyclotron motion a minimum in flight time is observed at
the frequency of the unperturbed cyclotron frequency. Upon variation
of the coupling frequency a resonance line is obtained whose shape
can be calculated \cite{Koenig1995}, as shown in figure
\ref{fig:cyclotron-resonance}. The full width at half minimum is
given by $\Delta \nu_{1/2} = 1 / T_{rf}$.

\section{Experimental facilities}

\begin{figure} [htbp]
    \centering
        \includegraphics[width=0.9\textwidth]{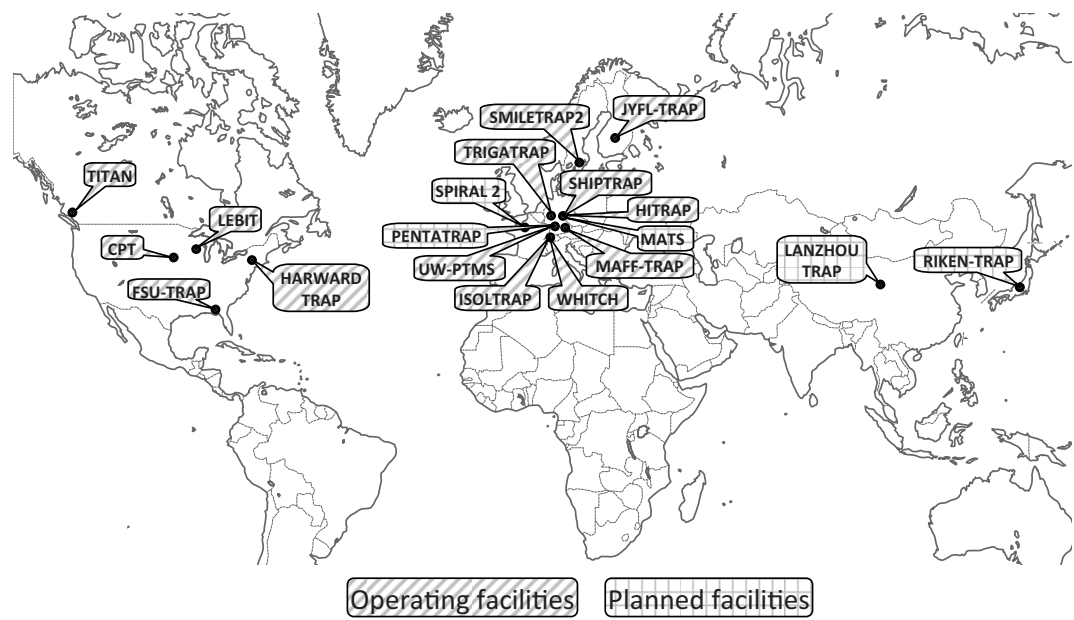}
    \caption{Existing and planned experiments for Penning trap mass determination of exotic ions.}
    \label{fig:facilities}
\end{figure}

A number of trap experiments worldwide, as shown in figure
\ref{fig:facilities}, have been constructed or are planned to
determine masses of exotic species, i.e. short-lived radionuclides
or highly charged ions. These experiments to a large degree are
complementary, they differ mainly by the way the ions are produced
and transported to the trap at the corresponding facilities
\cite{Bollen2004,Aysto2008}. Some of them are designed to work with
extremely short-lived isotopes. For example, at the TITAN experiment
at TRIUMF recently a measurement was carried out on $^{11}$Li with a
half life of $t_{1/2}=8.8$\,ms, so far the shortest lived nuclide
for which a mass measurement has ever been performed with a Penning
trap \cite{Smith2008}. The SHIPTRAP experiment at GSI is dedicated
to high-precision mass measurements of trans-uranium elements and
rare isotopes produced in fusion-evaporation reactions where the so
far heaviest element, $^{254}$No, has been measured in a Penning
trap \cite{Block2005}. The LEBIT experiment at NSCL/MSU is
exploiting short-lived rare isotopes produced by fast-beam
fragmentation \cite{Ringle2006}. JYFLTRAP at the IGISOL facility of
the University of Jyv\"askyl\"a focuses on exotic radioactive ion
beams of refractory elements, neutron-rich nuclides and
super-allowed beta emitters \cite{Jokinen2006}. The CPT mass spectrometer is investigating 
neutron-deficient nuclei produced in fusion-evaporation reactions using beams from the 
Argonne Tandem Linear Accelerator System (ATLAS) \cite{Clark2003}. The recently
commissioned TRIGA-TRAP experiment at the research reactor TRIGA
Mainz will perform high-precision mass measurements on thermal
neutron induced fission products, exotic neutron-rich nuclides and
actinides \cite{Ketelaer2008}. Highly charged ions are used in some
facilities, e.g. SMILETRAP in Stockholm, where the so far highest
charge state measured was $^{204}$Hg$^{52+}$ \cite{Fritioff2006}.
Using highly charged ions increases the precision of the mass
determination since the cyclotron frequency scales linearly with the
charge state, whereas the full width of the resonance line is
determined only by the duration of the excitation $T_{{rf}}$.

\begin{figure} [htbp]
    \centering
        \includegraphics[width=0.80\textwidth]{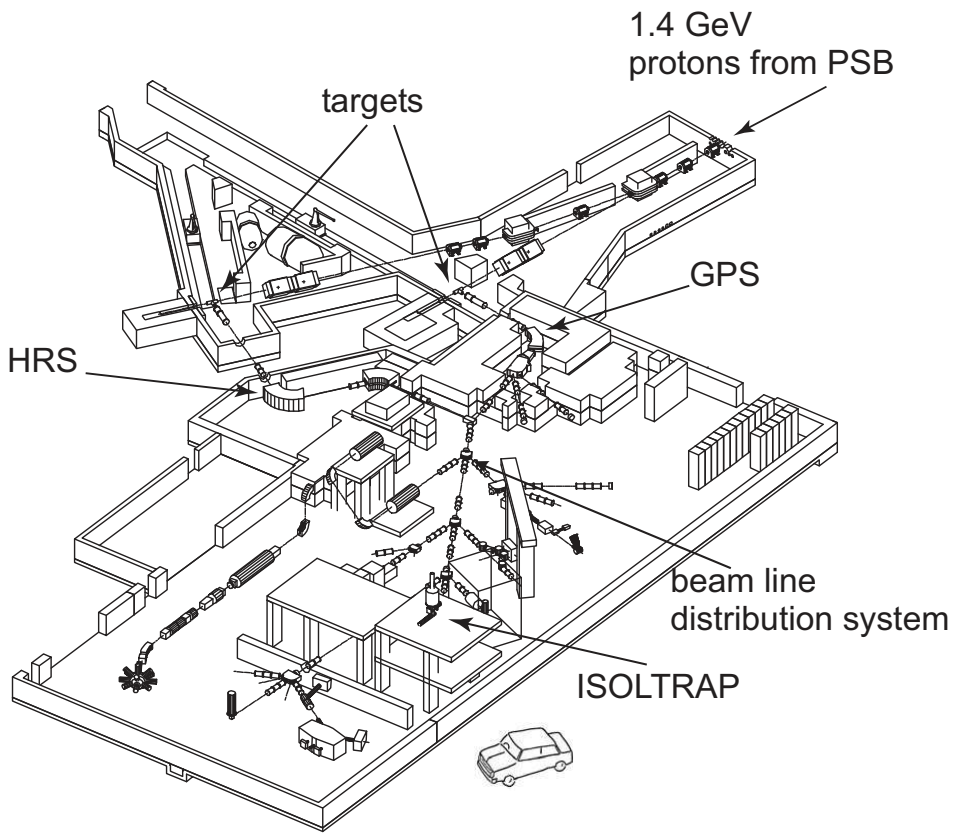}
    \caption{General layout of the CERN/ISOLDE facility. The new hall extension is not shown. For more details see text.}
    \label{fig:ISOLDE}
\end{figure}

\begin{figure} [htbp]
    \centering
        \includegraphics[width=0.80\textwidth]{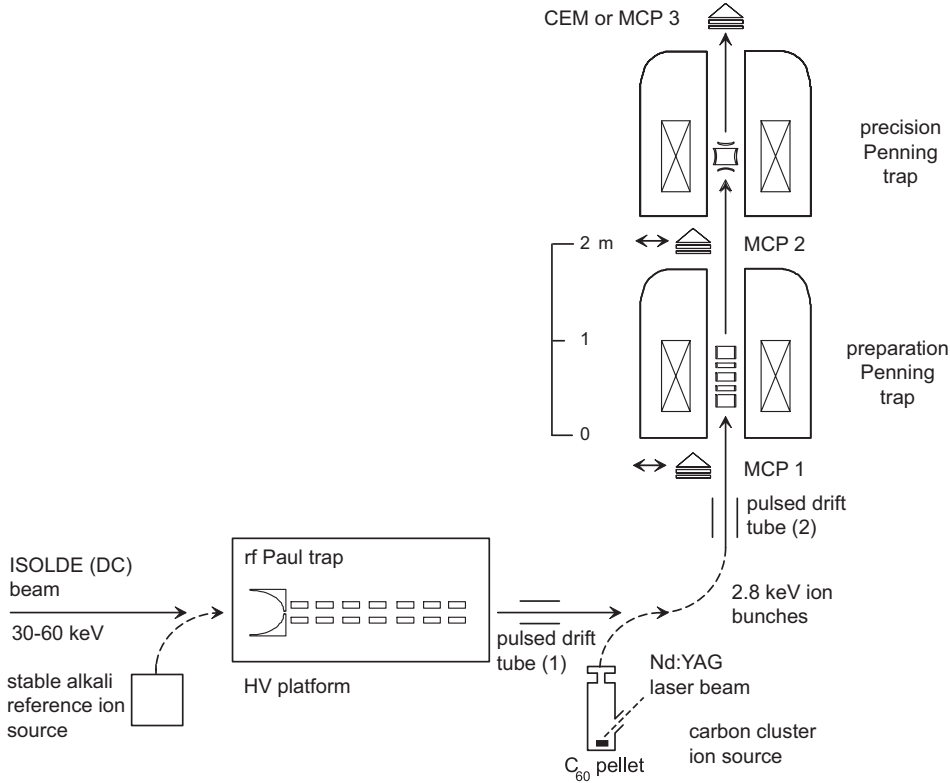}
    \caption{Sequence of traps for high-resolution mass spectrometry
    with ISOLTRAP at ISOLDE/CERN.}
    \label{fig:trap-sequence}
\end{figure}

Historically the first one of these on-line facilities is ISOLTRAP at
ISOLDE/CERN, which was initiated for more than two decades ago
\cite{Bollen1996,Mukherjee2008} and reaches relative mass
uncertainties of $\delta m/m < 10^{-8}$ \cite{Blaum2003c}. It has
produced a vast number of results on unstable isotopes and may serve
as typical example for a Penning trap mass spectrometer experiment,
thus we shall briefly describe its main features. Figure
\ref{fig:ISOLDE} shows the general layout of the ISOLDE facility. The
unstable nuclides are produced by nuclear spallation, fission, or
fragmentation reactions induced by 1.4-GeV protons arriving at the
target station of either the General Purpose Separator (GPS) with a
mass resolving power of $m/\Delta m \approx 1000$ or at the High
Resolution Separator (HRS) with $m/\Delta m \approx 5000$
\cite{Kugler2000}. After diffusion the radionuclides are ionized by
surface ionization, plasma discharge, or resonant laser ionization,
depending on the species of interest. The beam is then guided at
30 to 60\,keV to a linear radiofrequency quadrupole (RFQ) trap filled
with buffer gas \cite{Herfurth2001}. It acts to bunch and cool the
incoming continuous high energy beam by collisions with the buffer
gas. After a time of several ms a bunch of ions is extracted. These
bunches can be efficiently transported and captured in the first
cylindrical Penning trap (\emph{preparation trap} in figure
\ref{fig:trap-sequence}), where a mass-selective buffer gas cooling
is performed. This involves that the magnetron radius of all ion
species regardless of their mass is increased to a larger diameter
than the size of the exit aperture of the trap by a dipolar
excitation at the magnetron frequency. A mass selective quadrupolar
excitation on the cyclotron frequency centers the ion of interest in
the presence of the buffer gas. This leads to ion purification since
only the centered ions of reduced radius can leave the trap while
other ions with large magnetron radius are held back. The mass
resolution of this purification step is about $10^{5}$. Finally, the
ions are re-captured in the second trap (\emph{precision trap} in
figure \ref{fig:trap-sequence}), where the mass measurement takes
place. The cyclotron resonance is detected by the time-of-flight
technique described previously. The mass resolution depends on the
excitation time and is typically of the order of $10^6$--$10^{7}$,
allowing to resolve even low-lying isomeric states
\cite{Blaum2004b,VanRoosbroeck2004,Weber05}. Alternating with the
ions of interest stable reference ions like alkaline ions from a
surface ion source or carbon cluster ions produced by laser ablation
from a carbon pellet are loaded into the trap. Their cyclotron
frequency serves for calibration of the magnetic field
\cite{Kell2003,Blaum2003e}.

\section {Selected applications}

Here we present a number of examples which emphasizes that high
precision mass measurements contribute significantly to various
fields of physics.

\subsection{Nuclear structure studies and test of the isobaric-multiplet mass equation}

The mass of a nucleus is less than the sum of the masses of the
individual protons and neutrons. The difference between the sum of
the masses of $Z$ protons $Zm_{{p}}$ and $N$ neutrons $Nm_{{n}}$ and
the measured mass of the bound system $m$ is the nuclear binding
energy $B$:

\begin{equation}
B(N,Z) = (Zm_{{p}}+ Nm_{{n}} - m(Z,N)) \ c^{2}.
\label{eq:binding-energy}
\end{equation}

\noindent It depends on details of the nuclear composition. Mass
measurements with fractional uncertainties below $10^{-6}$ allow
discussing the hills and valleys of the binding energy appearing
across the nuclear chart in view of nuclear models
\cite{Lunney2003,Schwarz01,Hager2007,Guenaut2007,Hakala2008,Weber2008}. 

As a specific case we take the case of light nuclei with isobaric analog
states. They have nearly identical wave functions and the charge
dependent energy difference of these states can be calculated by
first order perturbation theory assuming only two-body Coulomb
interaction. This leads to the \emph{Isobaric Multiplet Mass
Equation} (IMME) which gives the mass $m$ of a member of an isospin
multiplet as function of the isospin $z$-component $T_{{z}} =
(N-Z)/2$ \cite{Benenson1979,Britz1998}:

\begin{equation}
M = a + bT_{{z}} + cT_{{z}}^{2}.
\label{eq:IMME}
\end{equation}

\begin{figure} [htbp]
    \centering
        \includegraphics[width=0.75\textwidth]{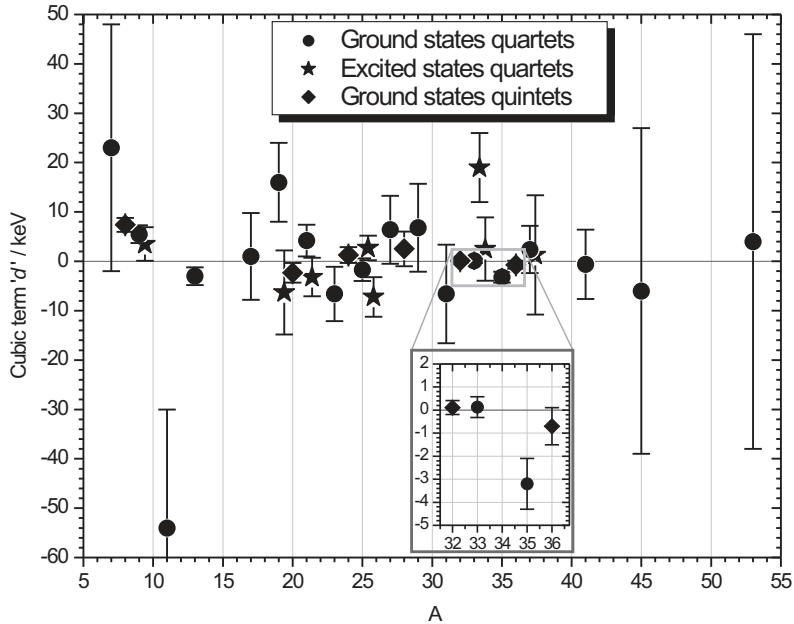}
    \caption{Size of an assumed cubic term in the isobaric-multiplet
    mass equation IMME. The data points in the enlarged inset are based on mass measurements of Ar ($A=32, 33$)
    and K ($A=35, 36$) isotopes and represent the most stringent tests performed so far
    on the quadratic form of the IMME \cite{Blaum2003b,Yazidjian2007}.}
    \label{fig:cubic-term}
\end{figure}

\noindent It is generally assumed that the quadratic term in this
equation is adequate. The so far most stringent tests of this
assumption have been performed at ISOLTRAP using mass measurements
on $^{32}$Ar ($T_{1/2}$ = 98\,ms), $^{33}$Ar ($T_{1/2}$ = 173\,ms),
$^{35}$K ($T_{1/2}$ = 178\,ms), and $^{36}$K ($T_{1/2}$ = 342\,ms)
with fractional uncertainties of a few parts in $10^{-8}$ together
with mass excess values of the other states of the $T = 2$ quintet
in $A = 32$ and $A = 36$, respectively, and the $T = 3/2$ quartet in
$A = 33$ and $A = 35$, respectively \cite{Blaum2003b,Yazidjian2007}.
An assumed cubic $dT_{{z}}^{3}$ term in equation (\ref{eq:IMME}) was
found to be consistent with zero within the error bars for three out
of four cases, as shown in figure \ref{fig:cubic-term}. Only for the
$A = 35$, $T = 3/2$ case a non-zero $d$ coefficient of
$-3.2(1.1)$\,keV was found, for which the reason is still unclear
\cite{Yazidjian2007}.

\subsection{Nuclear astrophysics studies}

One of the problems in nuclear astrophysics is a detailed
understanding of the formation of matter in stars. Three main 
scenarios are presently considered: Slow neutron capture
(s-process) is responsible for the creation of nuclei close to the
valley of stability.  Rapid neutron capture (r-process) causes the
formation of heavy nuclei such as uranium and thorium. Finally, rapid
proton capture (rp-process) creates nuclei close to the proton drip
line. The s-process is believed to be well understood, primarily
because the masses and the lifetimes of the involved nuclei are well
known. Concerning the rp-process (or the recently discussed
$\nu$p-process) many masses of interest have already been addressed,
see e.g. \cite{Fallis2008,Weber2008b,Weber2008c}. However, many open
questions appear in our understanding of the r-process since masses
of isotopes containing about 30 neutrons more than in the heaviest
measured isotope of the same element are unknown. Presently these
masses are not accessible to Penning trap mass measurements by
existing radioactive beam facilities. Therefore nuclear astrophysics
relies on nuclear mass models and predictions. Improvement of the
models and the reliability of the predictions require the
measurement of masses as far away of the stable area as possible.
Most important is the region where the r-process comes closest to
stability. This is the case in the vicinity of magic neutron
numbers. Crucial candidates are, e.g., $^{78}$Ni, $^{80}$Zn,
$^{132}$Sn, and $^{208}$Pb. Two recent experiments at ISOLDE have
determined the masses of $^{80,81}$Zn \cite{Baruah2008} and
$^{132,134}$Sn \cite{Dworschak2008}, respectively, with uncertainties of $10^{-7}$
and below and represent cornerstones in our understanding of the
r-process.

\subsection{$^{3}H$ - $^{3}He$ mass difference}

The $\beta$-decay of $^{3}$H
\begin{equation}
{^{3}H \rightarrow \ ^{3}He + e^{-} + \bar{\nu}}
\label{eq:mass-difference}
\end{equation}
\noindent is used for attempts to determine the mass of the electron
(anti)neutrino. Due to a low mass difference, i.e. $Q$-value, and
relatively short half life ($t_{1/2}$\,=\,12.32 years), the $^3H$ decay
reaction is rather advantageous. Such experiment involves the
measurement of the decay spectrum of tritium at the highest energies
near the endpoint energy, which is given by the mass difference
between $^{3}$H and $^{3}$He. A finite neutrino mass would change
the shape of the spectrum. The spectrum has been measured and
presently the upper limit of the electron neutrino mass is
$m_{\nu_e} < 2$\,${eV/c^{2}}$ at 95$\%$ C.L. \cite{Otten2008}.

The $^{3}$H - $^{3}$He mass difference is an input parameter in the
fitting process of the decay spectrum. It has been determined in a
Penning trap already in 1993 \cite{Vandyck1993}, and today's best
value recently measured at SMILETRAP is 18\,589.8(1.2)\,eV \cite{Nagy06}.  
A new $\beta$-spectroscopy experiment, KATRIN, is presently being setup at Forschungszentrum Karlsruhe, Germany, and aims at lowering the $m^2(\nu_e)$ uncertainty by
another factor of 100, thus reaching a sensitivity limit $m_{\nu_e}
< 0.2$\,${eV/c^{2}}$ \cite{Otten2008,Bornschein2005}. Consequently
measurements of the $^{3}$H and $^{3}$He masses to below $10^{-11}$
are necessary to match the requirement of the KATRIN experiment.
Efforts in this direction are pursued at MPI-K Heidelberg in
collaboration with University of Washington, Seattle.

\subsection{Bound state QED tests}

The $g$-factor of the electron bound in simple atomic systems can be
calculated in the frame of bound-state quantum electrodynamics
(BS-QED). A comparison of experimental and theoretical values
represents a stringent test of the calculations \cite{Werth2006}.
Experiments have been performed on hydrogen-like $^{12}$C$^{5+}$
\cite{Haeffner2000} and $^{16}$O$^{7+}$ \cite{Verdu2004}. In a
Penning trap the electrons spin precession frequency $\omega_{{L}} =
g (qe/2m_e) B$ as well as the ions cyclotron frequency $\omega_{{c}}
= (qe/m) B$ have been measured to about 10$^{-9}$. The $g$-factor
follows as

\begin{equation}
g = \frac{\omega_{{L}}}{\omega_{{c}}} \cdot \frac{m_e}{m}.
\label{eq:g-factor}
\end{equation}

\noindent Obviously the value of the ions mass $m$ is required to
the same precision as the frequencies for $g$-factor determination.
This is the case for carbon and oxygen. The BS-QED contributions
scale approximately with the square of the nuclear charge $Z$ for
hydrogen-like ions. Experiments on systems with higher $Z$ would
therefore allow testing the BS-QED part to a higher degree
\cite{Vogel2008}. Experiments on Ca$^{19+}$ are under way
\cite{Schabinger2007} and higher-$Z$ ions are considered in the near
future at the GSI-HITRAP facility \cite{Kluge2008}. The
corresponding masses need to be determined accurately. In the case
of $^{40}$Ca for example this has been already performed at
SMILETRAP with an uncertainty of $6 \cdot 10^{-10}$ \cite{Nagy2006}.

\subsection{Fundamental constants}

Precise mass measurements provide possibilities to improve the value
of some fundamental constants or to find alternative ways of their
determination. One of these constant is the electron mass. It has
been determined several years ago by the comparison of the cyclotron
frequencies of electrons and carbon ions which serve as mass
reference \cite{Farnham1995}. Recently an alternative method has led
to a somewhat more precise value: The $g$-factor of the electron
bound in hydrogen-like ions has been determined on $^{12}$C$^{5+}$
\cite{Haeffner2000} and $^{16}$O$^{7+}$ \cite{Verdu2004} as
mentioned above. The theory of bound-state quantum electrodynamics
allows to calculate the $g$-factor to high precision
\cite{Pachucki2005}. If we take the calculated value for $g$ as
granted we obtain with the measured frequencies $\omega_{{L}}$ and
$\omega_{{c}}$ from equation (\ref{eq:g-factor}) a value for the
electron/ion mass ratio. Since the masses of carbon and oxygen are
known with high precision it follows a value for the electron mass.
The combined value from both experiments is $m =
0.000\,548\,579\,909\,3(3)$ u \cite{Beier2002}. This may be improved
in the near future when experiments on low-$Z$ H-like ions are
performed \cite{Werth2006}. Here the BS-QED contributions as well as
other corrections from nuclear structure are small and their
uncertainty will not contribute significantly to the theoretical
$g$-factor. Potential candidates might be $^{24,26}$Mg$^{11+}$.
Their masses have been determined by the SMILETRAP Penning trap mass spectrometer with
relative uncertainties of $\delta m / m = 10^{-9}$ \cite{Bergstroem2003}.

Another important quantity is the fine structure constant $\alpha$
which defines the strength of the electromagnetic interaction. Its
most precise value comes from a comparison of the experimental and
theoretical values of the free electrons $g$-factor
\cite{Hanneke2008}: $\alpha^{-1} = 137.035\,999\,084\,(51)$ [0.37
ppb]. Since it relies on very difficult and complex calculations it
would be desirable to obtain a value for $\alpha$ independent of QED
theory. A possible way represents a combination of different
quantities which can be determined independently
\cite{Kinoshita1996}:
\begin{equation}
\alpha^{2} = \frac{2R_{{\infty}}}{c} \cdot \frac{h}{m_{{Cs}}} \cdot \frac{m_{{Cs}}}{m_{{p}}} \cdot \frac{m_{{p}}}{m_{{e}}}.
\label{eq:fine-structure-constant}
\end{equation}
The Rydberg constant $R_{{\infty}}$ is known to $6.6 \cdot 10^{-12}$ \cite{Codata2006},
$h/m_{{Cs}}$ has been determined by photon recoil to $7.3 \cdot 10^{-9}$
\cite{Wicht2002}. The last part in equation
(\ref{eq:fine-structure-constant}) relates to measurements of mass
ratios $m_{{Cs}}/m_{{p}}$ and $m_{{p}}/m_{{e}}$. Future improvements
of all quantities appearing in equation
(\ref{eq:fine-structure-constant}) may lead to an independent
determination of $\alpha$ with somewhat comparable accuracy as the
QED related value \cite{Chu2006,Biraben2006,Chu2008}.

\subsection{Neutrino-less double $\beta$-decay}

In double $\beta$-decay ($\beta$$\beta$) processes two electrons are
emitted from a nucleus and the charge increases by 2. While the
simultaneous emission of two neutrinos is an allowed second order
process of weak interaction, neutrinoless double $\beta$-decay can
only occur if neutrinos are massive Majorana particles
\cite{Vergados2002}. It would represent a violation of the classical
Standard Model. In order to set a limit on the neutrinoless decay
mode one needs to know the $Q$-values of the respective decays. The
presently best studied candidate is $^{76}$Ge. The $Q$-value for the
double $\beta$-decay has been determined by mass measurements of
$^{76}$Ge and $^{76}$Se in a Penning trap with uncertainties below
$10^{-9}$ \cite{Douysset2001}. The $Q$-value of 2039.006(50) deduced
from this experiment deviates by several standard deviations from
the value obtained from decay energy measurements
\cite{Klapdor2004}. This makes the report of an indication of an
observed neutrinoless double $\beta$-decay in \cite{Klapdor2004} and
the deduction of a finite mass for the electron neutrino between 170
and 630\,meV somewhat questionable. Other $Q$-values of double-beta
decays that have been addressed recently by high-precision Penning
trap mass spectrometry are the ones of $^{100}$Mo \cite{Rahaman2008}
and $^{136}$Xe \cite{Myers2007}.

\section{Future developments}

The challenge for future high-precision mass measurements is to
extend the range of presently available isotopes towards shorter
lived ones, further away from the region of stability, and to reach
higher mass resolving power and mass accuracy. Technical
improvements are needed towards this goal. These include using new
excitation schemes of the ion motion in the trap like Ramsey
excitation \cite{George2007,Kretzschmar2007,George2007b} or
octupolar excitation \cite{Ringle2007,Eliseev2007}, higher charge
states by charge breeding processes \cite{Dilling2006,Boehm2005,Crespo2004}, laser
cooling of ions, sympathetic ion cooling by simultaneously trapped
cold electrons \cite{Kluge2008}, and trap operation at cryogenic
temperatures \cite{Ketelaer2008,Ferrer2007}. Some of these
improvements are already implemented in operational facilities,
others are being considered for new facilities under construction.
The problem remains to get access to those isotopes whose production
rate is too small for present mass spectrometers. New facilities
coming up such as EURISOL/France \cite{URLeurisol}, FAIR/Germany
\cite{URLfair}, FRIB/USA \cite{URLria} and RIKEN/Japan
\cite{URLriken} will use improved techniques for ion sources,
synchrotrons, fragment separators and storage rings in order to
enhance the intensity of radioactive beams by several orders of
magnitudes. This will substantially widen the range of accessible
nuclides and will allow future experiments of fundamental interest.

\ack Financial support by the German Federal Ministry for Education
and Research (BMBF) under contract 06MZ215 and by the Helmholtz
Association for National Research Centers (HGF) under contract
VH-NG-037 is acknowledged. Sz. Nagy acknowledges the support of 
the Alliance Program of the Helmholtz Association.



\end{document}